\documentclass[12pt]{article}
\usepackage{titlesec}
\usepackage{epsfig}
\usepackage{graphicx}
\usepackage{graphics}
\usepackage{amssymb}
\usepackage{amsmath}
\usepackage{amsfonts}
\usepackage{color}
\usepackage{textcomp}
\usepackage{latexsym}
\usepackage{cite}
\usepackage{titletoc}
\usepackage{tikz}
\usepackage{xcolor}
\usetikzlibrary{3d}
\usepackage{csquotes}
\usepackage{caption}
\usetikzlibrary{calc} 
\usetikzlibrary{angles,quotes}
\usepackage{csquotes}
\titleformat{\section}{\normalfont\Large\bfseries}{\thesection.}{1em}{}
\titlecontents{section}[0em]{\bfseries}{\thecontentslabel. }{}{\titlerule*[1pc]{.}\contentspage}
\usepackage[margin=2.5cm,footskip=0.6cm]{geometry}
\begin{document}

\begin{center}
{\Large{\bf Ultraslow growth of domains in a random-field system with correlated disorder }} \\
\ \\
\ \\
by \\
Subhanker Howlader and Prasenjit Das\footnote{prasenjit.das@iisermohali.ac.in}\\
\textit{Department of Physical Sciences, Indian Institute of Science Education and Research Mohali, Knowledge City, Sector 81, SAS Nagar, Mohali, Punjab 140306, INDIA}\\
\vspace{0.5cm}
Manoj Kumar\footnote{manoj\_kumar@niser.ac.in} \\
\textit{School of Physical Sciences, National Institute of Science Education and Research Bhubaneswar, Jatni 752050, India\\  and Homi Bhabha National Institute, Training School Complex, Anushakti Nagar, Mumbai 400094, India}
\end{center}

\begin{abstract}
\noindent 
We study domain growth kinetics in a random-field system in the presence of a spatially correlated disorder  $h_{i}(\vec r)$ after an instantaneous quench at a finite temperature $T$ from a random initial state corresponding to $T=\infty$. The correlated disorder field $h_{i}(\vec r)$ arises due to the presence of magnetic impurities, decaying spatially in a power-law fashion. We use Glauber spin-flip dynamics to simulate the kinetics at the microscopic level. The system evolves via the formation of ordered magnetic domains. We characterize the morphology of domains using the equal-time correlation function $C(r,t)$ and structure factor $S(k,t)$. In the large-$k$ limit, $S(k, t)$ obeys Porod's law: $S(k, t)\sim k^{-(d+1)}$. The average domain size $L(t)$ asymptotically follows \textit{double logarithmic growth behavior}.
\end{abstract}

\newpage
\section{Introduction}\label{sec1}
Consider a magnetic system of spin-$\frac{1}{2}$ atoms that resides in a paramagnetic state at a temperature $T$, above the critical temperature $T_c$ for the paramagnetic-to-ferromagnetic transition. When this system is rapidly quenched to $T<T_c$, it falls out of thermodynamic equilibrium. As a result, the system evolves by forming domains enriched with spin-up $s_i = +1$ or spin-down $s_i = -1$ atoms~\cite{dp04,pw09}. In absence of disorder or impurities (pure systems), average domain size $L(t)$ grows with time as $L(t)\sim t^{1/2}$, which corresponds to diffusive growth with a non-conserved order parameter, known as Lifshitz-Allen-Cahn (LAC) law~\cite{pw09,ac79,b94}. This is observed when the domains grow without any energy barrier ($E_{\rm B}$), or when $E_{\rm B}\ne 0$ but remains independent of $L(t)$~\cite{p04}. The domain morphology can be studied by studying the equal-time order-parameter correlation function $C(r, t)$, and its Fourier transform, the structure factor $S(k, t)$~\cite{pw09,ddp17}. If the domain growth process is governed by a single characteristic length scale, both $C(r, t)$ and $S(k, t)$ exhibit dynamical scaling with the scaling forms~\cite{b94,dp04,pw09}:
\begin{align}
  \label{eqn1}
  C(r,t) = &g[r/L(t)],\\
  \label{eqn2}
  S(k,t) = &L(t)^{d}f[kL(t)].
\end{align}
Here, $g(x)$ and $f(p)$ are time independent master functions, and $d$ is the spatial dimensionality. The functional form of $g(x)$ is given by the well-known \textit{Ohta-Jasnow-Kawasaki}~(OJK) function~\cite{dp04,pw09}.

In real systems, nonmagnetic impurities, vacancies, and a local random field on lattice sites are the common form of disorder, which arise due to structural imperfections. These factors introduce heterogeneity in the domain growth process. As a result, domain walls get pinned around disordered regions, leading to a slowing down of domain growth in the late stage of evolution~\cite{optdc02}. At the microscopic level, the Ising model with nearest-neighbor~(NN) interactions serves as a fundamental framework for describing magnetic systems~\cite{dp04,pw09}, which can be used to model the real systems of disorder with the Hamiltonian given by
\begin{align}
  \label{eqn3}
  \mathcal{H} = -\sum_{\langle ij\rangle}J_{ij}s_is_j - \sum_i h_i s_i,
\end{align}
where $s_i=\pm1$ and $J_{ij}$ is the random interaction strength between spins $s_i$ and $s_j$. The summation $\langle ij\rangle$ runs over all nearest-neighbor sites. $h_i$ are the local random-field variables, which are usually drawn from a certain probability distribution. For pure systems, we consider $J_{ij}=J$ and $h_i=h$.  Disorder is typically introduced into the system in two primary ways:  either by varying the strength of the exchange interactions between neighboring spins $J_{ij}$ or by incorporating a local random field $h_i$ into the Ising Hamiltonian. The former is known as \textit{random-exchange Ising model} (REIM)~\cite{dp04,p04} while the latter is well-known as \textit{random-field Ising model} (RFIM)~\cite{n98,dp04,p04}. The equilibrium properties and non-equilibrium dynamics of the REIM~\cite{h74,hl74,gl76,pc91,bh91,bpc96,pr04,pp05,ps07,lmp10,pp12,cz15,cc15} and RFIM~\cite{n98,imo75,gg84,a87,nv88,nr89,oc90,pp93,rc293,o94,accp08,CLMPZ12,ms14} have been the subject of extensive research over the years. Below, we summarize some key findings from previous studies in disordered ferromagnets in the context of both the REIM and RFIM, which are relevant to our present study.

The foundational study on ordering in ferromagnetic REIM was due to Huse and Henley~\cite{hh85}. They observed that bond disorder causes domain walls to become pinned at local free energy minima. As a result, the movement of domain walls, which governs the process of domain growth, is linked to the crossing of energy barriers $E_{\rm B}$, resulting in a logarithmic asymptotic growth law: $L(t)\sim (\ln t)^x$. The exponent $x$ depends on how $E_{\rm B}$ depends on $L$. This was further supported by theoretical and numerical work from Puri \textit{et al.}~\cite{pp92} and experimental findings by Schins \textit{et al.}~\cite{sa93}, who observed a similar growth law. Several research groups have characterized the morphology of domains by calculating $C(r, t)$ and $S(k, t)$ of the order parameter field, which exhibit dynamical scaling. Additionally, they have explored the concept of superuniversality~(SU), which suggests that scaling properties are independent of specific microscopic details and apply across a range of disordered systems~\cite{bh91,pp12,hm06,hp08,kc20}. Further, systems such as dilute magnets~\cite{cz15, clm13} and spin glasses~\cite{by86,h11} can be described by REIM with certain fractions of bonds having $J_{ij}=0$ for dilute magnets and $J_{ij}<0$ for spin glasses. Rammal and Benoit studied critical dynamics in the dilute Ising model~\cite{rb85}, identifying a distinct asymptotic growth law near the percolation threshold, given by $L(t)\sim t^\alpha$, where $\alpha$ depends on $T$ and disorder amplitude. The comprehensive investigation of Paul \textit{et al.}~\cite{pr04,pp05} confirmed this domain growth law, aligned with experimental observations by Shenoy\textit{et al.}~\cite{ss99} and Likodimos \textit{et al.}~\cite{ll00,llo01}. 

For  both models RFIM and REIM, the domain growth behavior at a finite $T$ is influenced by the presence of energy barriers~\cite{v84,gf84,gm83,ba84,aj85}. Previously, the kinetics of domain growth were studied by several groups using both microscopic approaches~\cite{CLMPZ12,gkg84,gkg85,pf85,ms14} and coarse-grained simulations~\cite{dp04,pp92,accp08,kc20,djpe20,ddpc24}. For systems with a nonconserved order parameter, the early evolution of $L(t)$ follows the LAC law: $L(t) \sim t^{1/2}$. As time advances, this growth law crosses over to a logarithmic form, $L(t) \sim (\ln t)^{\phi}$, consistent with experimental findings~\cite{sh95,fbh95,zw88}. However, there is a lack of complete understanding about the growth exponent $\phi$. The scaled correlation function $C(r,t)$ and the structure factor $S(k,t)$ exhibit dynamical scaling. Unlike pure systems where $C(r,t)$ decays linearly from its maximum value as $[1 - C(r,t) = br + \cdots]$, here it follows a cusp-like decay: $[1 - C(r,t) = ar^\theta + \cdots]$, where $\theta$ is the disorder-dependent cusp exponent~\cite{bs84,db00b08,skb11,CLMPZ12,sk14,kbp17}. The linear decay is indicative of sharp domain interfaces and leads to the Porod's law in the structure factor: $S(k,t) \sim k^{-(d+1)}$, where $d$ denotes the spatial dimensionality. In contrast, the cusp behavior in $C(r,t)$ leads to a fractal Porod law: $S(k,t) \sim k^{-(d+\theta)}$, where $\theta$ is related to the fractal dimension $d_m$ as $\theta = d_m-d$ for a mass fractal~\cite{n98,nv88,bpsij14}. The appearance of fractal Porod's law indicates the presence of fractal domains and interfaces. Additionally, analogous to REIM, the SU feature in RFIM has remained controversial as both SU violation \cite{CLMPZ12,kbp17,kc20} as well as SU preservation have been reported~\cite{accp08,rc293,bh91,hp08,skb11,sk14,kb17}.

In this paper, we investigate the ordering dynamics in a correlated random-field system quenched at a finite temperature. Unlike RFIM, where a random-field is assigned to each lattice point independently, our approach considers a scenario where only a fraction of lattice sites generate a nonlocal random-field, which acts as the centers of the disorder. These centres induce a local field at each site $i$; thereby introducing spatial correlations in the disorder. This leads to a distinct influence on the phase ordering process. Such conditions are achievable in systems doped with magnetite particles, which can produce magnetic fields due to their inherent magnetic properties, thereby creating spatial correlations in the disorder~\cite{rlzl07,ltbb23}. Our study aims to understand how these spatially correlated disorders on certain lattice sites affect the domain growth kinetics.

The structure of this paper is organized as follows. Sec.~\ref{sec2} outlines the details of our modeling approach. In Sec.~\ref{sec3}, we present the numerical methodologies and discuss the simulation results. Finally, Sec.~\ref{sec4} provides a summary of our findings and suggests potential avenues for future research.

\section{Details of Modeling}\label{sec2}
We examine the ordering dynamics on a two-dimensional (2D) square lattice in the presence of magnetic impurities. These impurities create a magnetic field around their locations that decays with distance according to a power-law. The magnetic field $h'(\vec r_i)$ at any site $\vec r_i$ due to an impurity at site $\vec r_j$  is described by the expression:
\begin{align}
  \label{eqn4}
  h'(\vec r_i) = \frac{h_0}{|\vec r_i - \vec r_j|^{d+\alpha}}.
\end{align}
where $h_0$ is the field strength, $d$ is the spatial dimensionality (here $d=2$) and the parameter $\alpha>0$ determines the decay of the field. To compute the total field  $h(\vec r_i)$ at site $\vec r_i$  due to impurity sites $\vec{r}_j$, we sum all contributions within a cutoff distance $|\vec{r}_i - \vec{r}_j| < r_0$, where $r_0$ is chosen significantly large enough so that the contributions to the total field beyond $r_0$ become negligible. Consequently, the field $h(\vec{r}_i)$ effectively behaves as a finite-range field.

We describe our system using the Ising Hamiltonian, as given by Eq.~(\ref{eqn3}). In this model, we set the strength of the NN exchange interaction to a constant, as $J_{ij}=J>0$. Without impurities, the critical temperature $T_c$ for the transition from the paramagnetic to the ferromagnetic phase is given by $T_c^{\rm pure}=2.269J/k_{\rm B}$, where $k_{\rm B}$ represents the Boltzmann constant. We introduce dynamics into this Ising model by coupling it to a heat bath that induces stochastic spin flips at lattice sites. This process is governed by the Glauber spin-flip model, where spins randomly flip in response to thermal fluctuations.

\section{Numerical Details and Results}\label{sec3}
We perform MC simulations on a square lattice of size $N=4096^2$. We employ periodic boundary conditions in all directions and set the cutoff distance $r_0=1024$ which is chosen to balance accuracy with computational efficiency. Any impurity site generates a magnetic field around it following Eq.~(\ref{eqn4}), effectively representing static or quenched impurities within the system. Next, we initialize spin variable $s_{i} =\pm 1$ randomly on all the lattice sites, which corresponds to a paramagnetic state at $T=\infty$. At time $t=0$, we rapidly quench the system to a finite $T=0.8$. After the quench, any stochastic move involves choosing a random site $i$. Once the site is selected, we virtually flip the spin $s_i$, resulting in a change in configuration from \{$s_i$\} $\rightarrow$ \{$s'_i$\}. The change is accepted with probability $p$, according to the Metropolis acceptance criteria~\cite{nb99,lb15}, given by:
\begin{equation}
\label{Eqn5}
p= \begin{cases} 
1 &; \Delta E \leq 0 \\ 
e^{-\beta \Delta E} &; \Delta E > 0 
\end{cases}
\end{equation}
Here, $\Delta E = \mathcal{H}(\{s'_{i}\}) - \mathcal{H}(\{s_{i}\})
$ is the energy difference between the final and initial configurations, $e^{-\beta \Delta E}$ is the Boltzmann weight factor, and $\beta=(k_{\rm B}T)^{-1}$ with $k_{\rm B}=1$. In the simulations, we fix the quench temperature $T=0.8$ and $\alpha=0.1$, and vary $\rho$ and $h_{\rm max}$. One Monte Carlo step (MCS) consists of $N$ attempted updates, which we also refer to as Monte Carlo time~(MCT), denoted by $t$. All the statistical results reported here are obtained as averages over ten independent runs of random selection of impurity sites as well as random initial conditions ($s_i=\pm1$).

\begin{figure}
    \centering
    \includegraphics[width=0.99\textwidth]{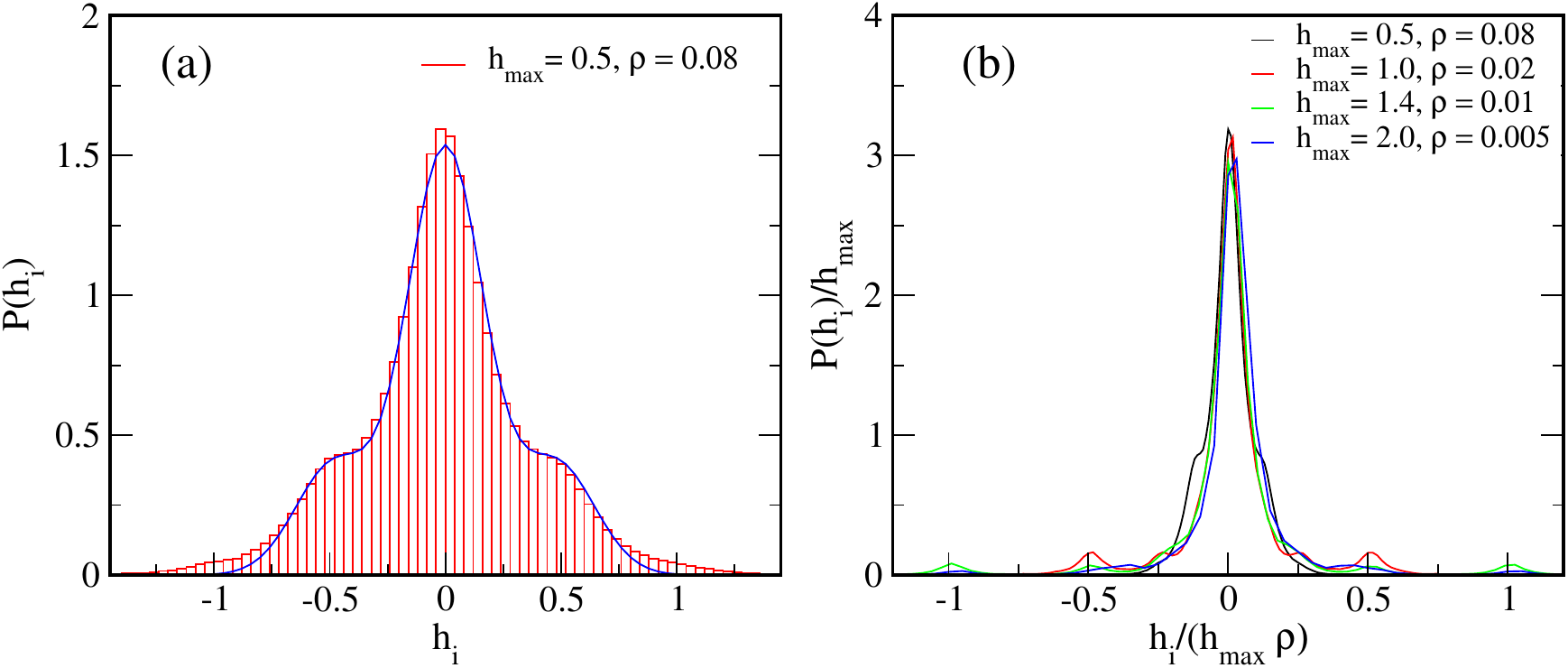}
    \caption{ (a) Distribution of the total field $h_i$ at site $i$ for $h_{\max}=0.5$ and $\rho=0.08$. The solid blue curve is the fit according to a central-dominant trimodal distribution. (b) Scaling of the  field distributions $h_i$ for different pairs of $h_{\max}$ and $\rho$, as marked.} 
    \label{h_i_dis}
\end{figure}

We begin by analyzing the distribution of the total fields $h_i$. Fig.~\ref{h_i_dis}(a) displays the distribution the total field $h_i$ for $h_{\max} =0.5$ and $\rho=0.08$, where a solid blue curve represents a fit to a central-dominant trimodal distribution of the form:
\begin{equation}
    P(h_i) = A_0 \exp\left[ -\frac{h_i^2}{2\sigma_0^2} \right]+A_1 \exp\left[ -\frac{\left(h_i-h_{\max}\right)^2}{2\sigma_1^2} \right]+A_1 \exp\left[ -\frac{\left(h_i + h_{\max}\right)^2}{2\sigma_1^2} \right],
\end{equation}
with $A_0>A_1$. The distribution clearly indicate the dominance of the central peak at $h_i=0$, which becomes even more pronounced for increasing $h_{\max}$ or lowering $\rho$. Figure~\ref{h_i_dis}(b) illustrates the scaling behavior of these distributions for various values of $h_{\max}$ and $\rho$. The distributions collapse with the scaling form:
\begin{equation}
    P(h_i)=h_{\max}~ \widetilde{p}\left(\frac{h_i}{h_{\rm\max}\rho}\right).
\end{equation}
From the normalization condition of $P(h_i)$,  it follows that
\begin{equation}
    \int \widetilde{p}(x) dx = 1/(h_{\max}^2\rho).
\end{equation}
This implies that the field amplitude $h_{\max}$ and density $\rho$ variables combine through a scaing variable $h_{\max}^2\rho$. In particular, we find that the width of the field distribution, quantified by the variance $\sigma^2 = \langle h_i^2\rangle-\langle h_i\rangle^2$ scales linearly with $h_{\max}^2\rho$, and is well approximated by 
\begin{equation}
 \sigma^2 \simeq 7 h_{\max}^2\rho.
 \label{sigma}
\end{equation}

\begin{figure}
    \centering
    \includegraphics[width=0.5\textwidth]{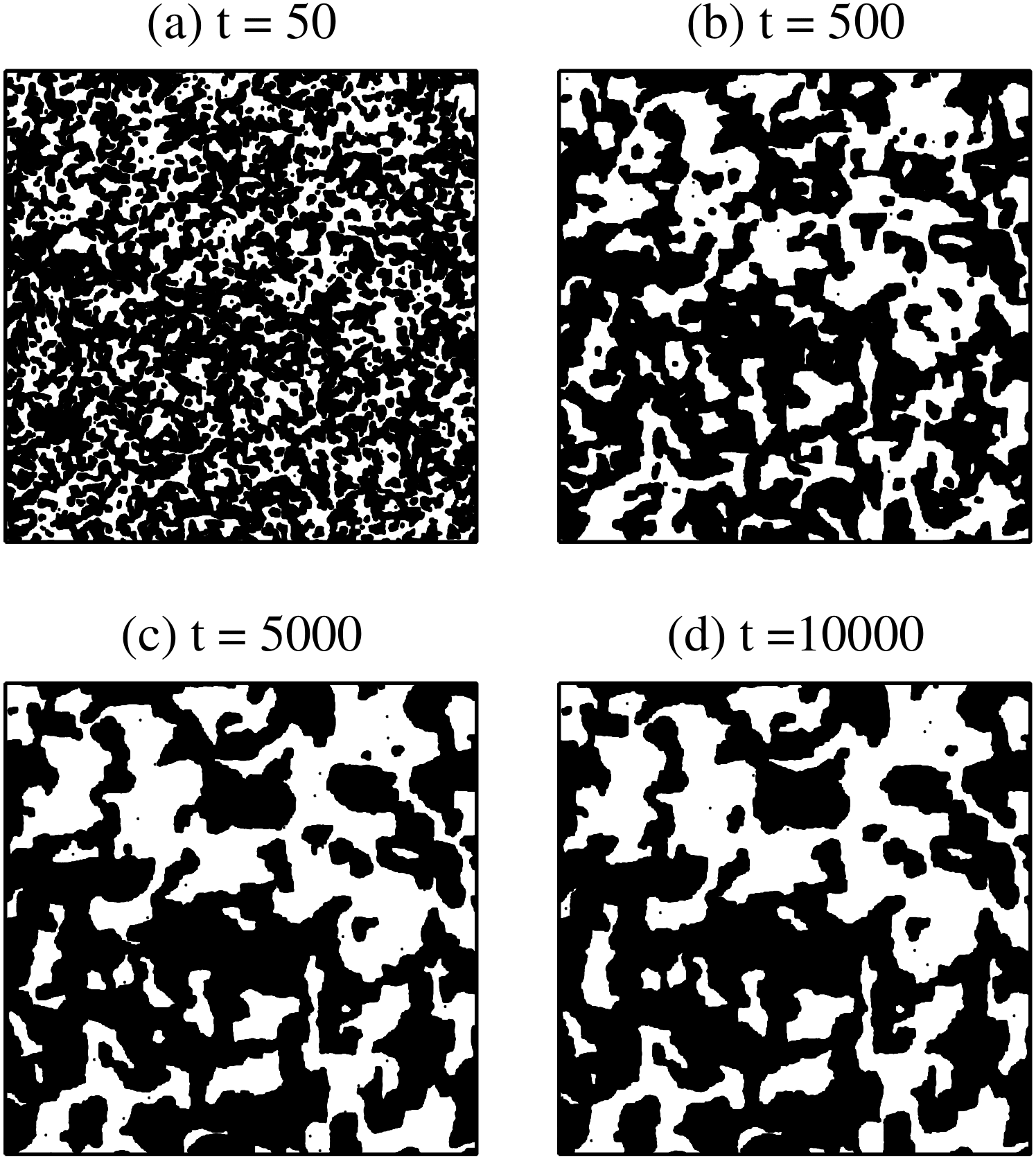}
    \caption{Snapshots showing the evolution of a domains at different Monte Carlo times $t$, as mentioned, after an instantaneous quench to $T=0.8$. Here, we set the maximum value of impurity-induced magnetic field $h_{\rm max}=0.5$ and the impurity density $\rho=0.01$. Lattice sites occupied by an up-spin are depicted in black, while sites with a down-spin are left unmarked. The actual system size is $4096\times 4096$ lattice sites, but for clarity, only a $1024\times 1024$ section from one corner is shown.}
    \label{Fig_1}
\end{figure}

Figure~\ref{Fig_1} shows the evolution snapshots of the spin configuration for $\rho=0.01$ and $h_{\rm max}=0.5$ at different MCTs $(t)$, as mentioned. As shown in the figure, we observe a clear formation of domains enriched with spins $s_i=+1$ and $s_i=-1$. As time advances, these domains grow in size by removing the interfaces between the differently ordered domains [cf. Fig.~\ref{Fig_1}(a) and Fig.~\ref{Fig_1}(b)]. However, this domain growth process does not persist for a long time. At the late stage of evolution, we see small structural changes in the domains, as shown in Fig.~\ref{Fig_1}(c) and Fig.~\ref{Fig_1}(d). This occurs due to the pinning of spins around the impurity sites, which creates local energy barriers. Consequently, domain growth predominantly happens through the thermally activated movement of domain walls over these barriers. However, at early times, when the average domain size is small, this effect is minimal. 

\begin{figure}
    \centering
    \includegraphics[width=0.5\textwidth]{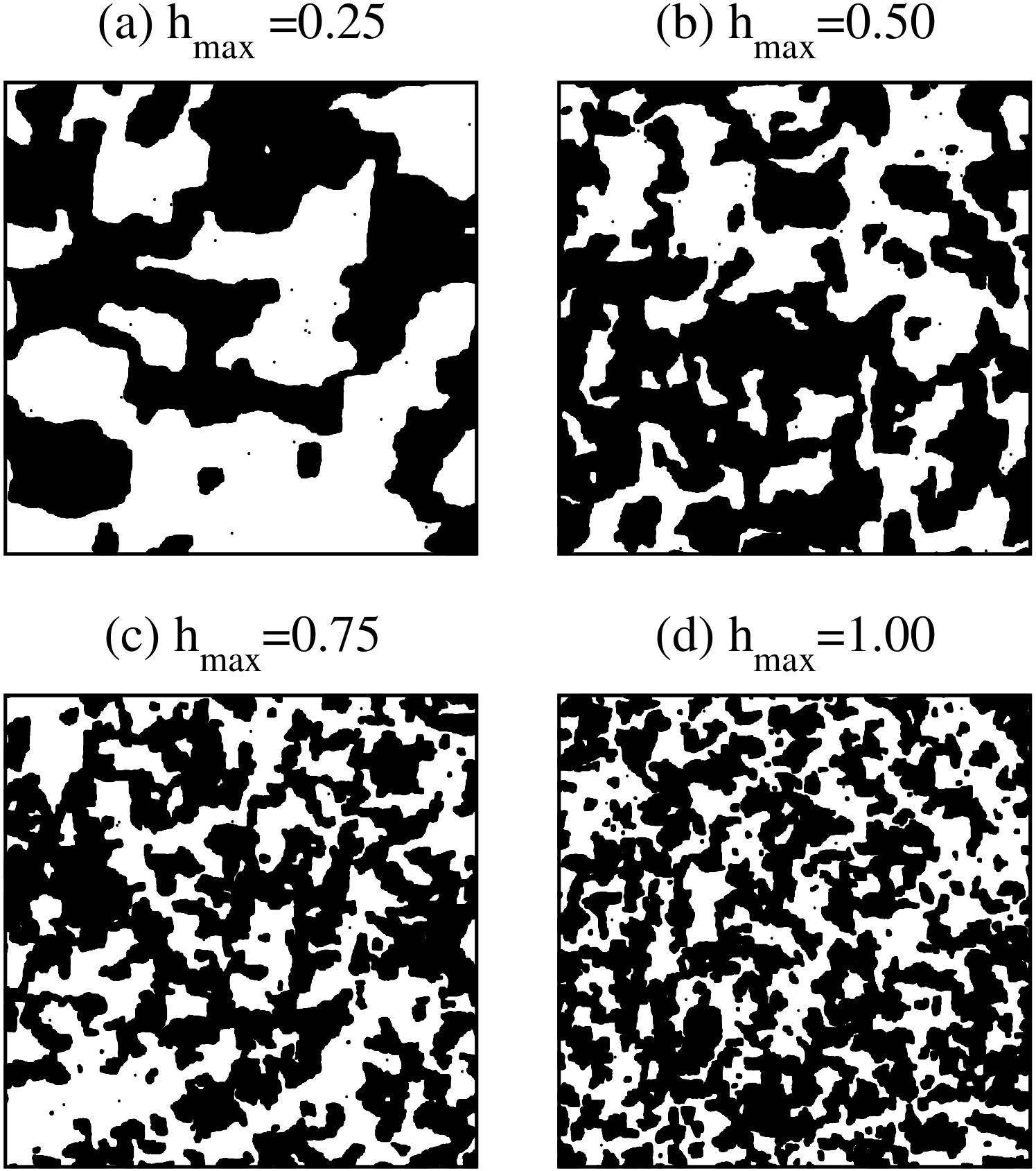}
    \caption{Snapshots at a fixed $t=6000$ for various values of $h_{\rm max}$, as mentioned. The impurity density is set to $\rho=0.01$. Rest of the details are the same as described in the caption of Fig.~\ref{Fig_1}.}
    \label{Fig_2}
\end{figure}

To explore the impact of the maximum field strength $h_{\rm max}$, on domain growth, we plot evolution snapshots of the system at a fixed  $t=6000$ for $\rho=0.01$ and for different values of $h_{\rm max}$ in Fig.~\ref{Fig_2}. Clearly, the average domain size depends on the value of $h_{\rm max}$. The higher the value of $h_{\rm max}$, the smaller the domain size. Since domain growth occurs through the crossing of energy barriers in the presence of thermal fluctuations at later times, and the barrier height increases with increasing $h_{\rm max}$, barrier jumps become less probable at a fixed $T$ for larger $h_{\rm max}$, thereby reducing the domain size.

\begin{figure}
    \centering
    \includegraphics[width=0.99\textwidth]{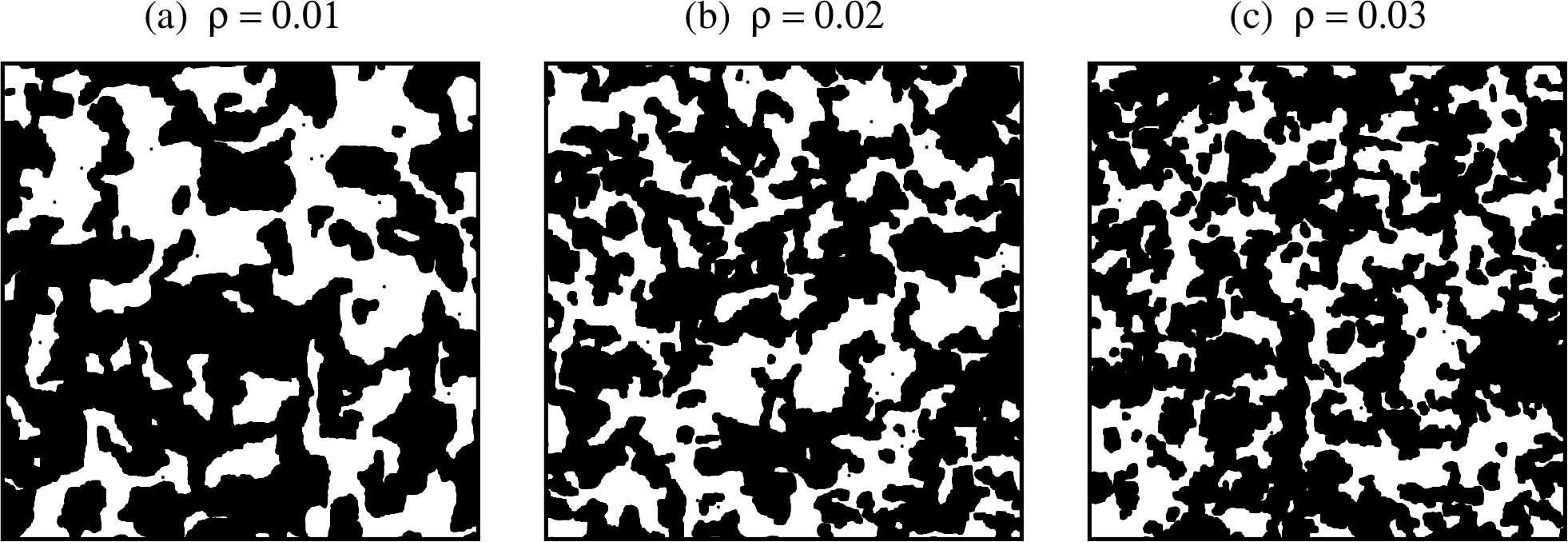}
    \caption{Snapshots of a disordered ferromagnet at $t=8000$ for various values of impurity density $\rho$, as mentioned. The maximum value of the impurity-induced field is set to $h_{\rm max}=0.5$. Further details are the same as mentioned in the caption of Fig.~\ref{Fig_1}.}
    \label{Fig_3}
\end{figure}

Next we examine the effect of the impurity concentration on the kinetics of domain growth at a fixed $h_{\rm max}$. Figure~\ref{Fig_3} shows the evolution snapshots of the system for fixed $h_{\rm max}=0.5$ and at $t=8000$ but for different values of $\rho$. It is evident that the average domain size decreases as $\rho$ increases. This is because the average distance between pairs of pinned lattice sites decreases with increasing $\rho$, making domain walls more susceptible to being trapped. Additionally, with higher $\rho$, we observe more pronounced trapping of domain interfaces.

\begin{figure}
    \centering
    \includegraphics[width=0.99\textwidth]{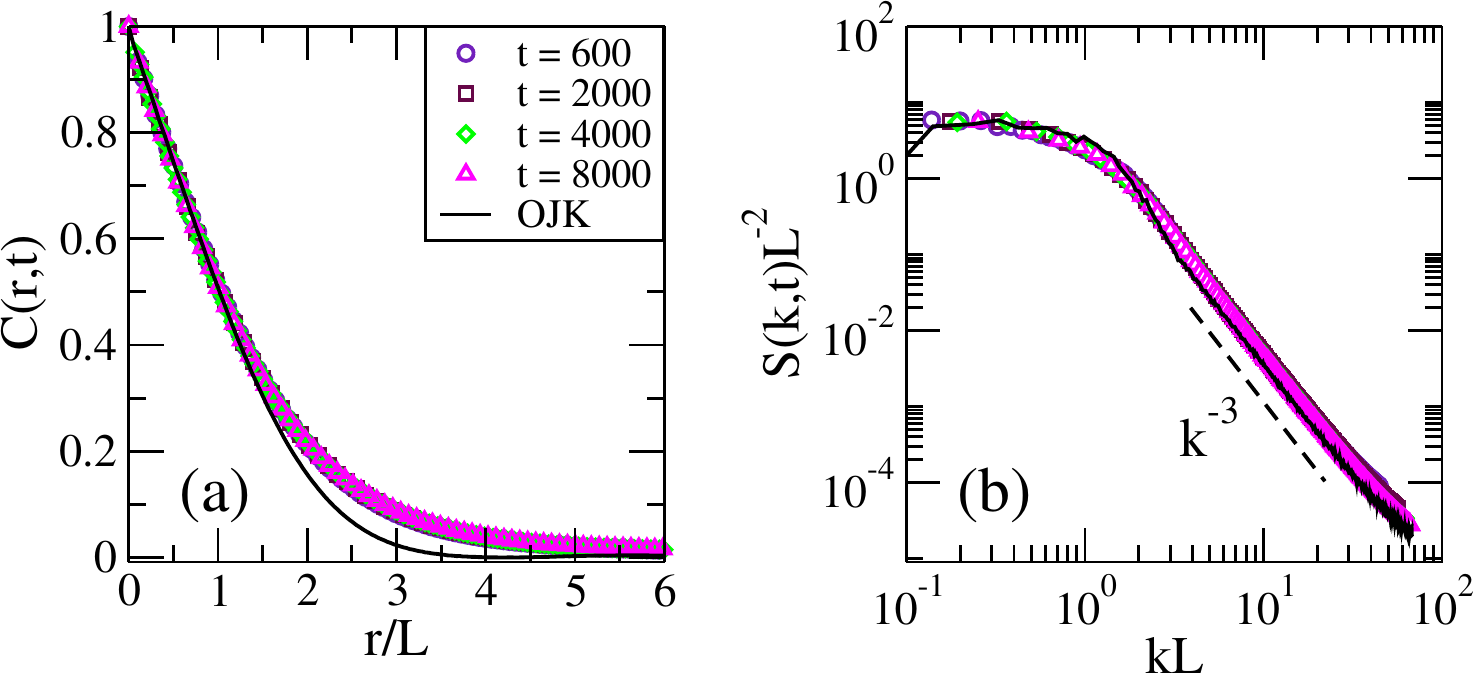}
    \caption{Scaling plot of correlation functions $C(r,t)$ and structure factor $S(k,t)$ for $\rho=0.01$ and $h_{\rm max}=0.5$. These functions are spherically averaged over ten independent MC runs. (a) Plot of $C(r,t)$ {\it vs.} $r/L$  at four different MCSs, as mentioned. The characteristic length scale $L(t)$ is defined as the distance at which $C(r,t)$ falls to half its maximum value, $1$ at $r=0$. The black solid line depicts the OJK function $C_{\rm OJK}(r, t)$. (b) Plot of $S(k,t)L^{-2}$ {\it vs.} $kL$ on a log-log scale. The symbols correspond to the same MCSs as in (a). The black solid line represents the Fourier transform of the OJK function. For large $k$, $S(k,t)$s follow Porod's law: $S(k,t)\sim k^{-(d+1)}$, with $d=2$, as indicated by the dashed line labeled $k^{-3}$.}
    \label{Fig_4}
\end{figure}

Next, we characterize the morphology of domains to understand the effect of $\rho$ and $h_{\rm max}$.
For that, we calculate the spherically averaged equal-time correlation function $C(r,t)$ and its Fourier transform structure factor $S(k, t)$. The correlation function is defined as follows:
\begin{align}
\label{Eqn6}
C(r, t)= &\langle s(\vec{R}, t)s(\vec{R}+\vec{r}, t)\rangle-\langle s(\vec{R}, t)\rangle\langle s(\vec{R}+\vec{r}, t)\rangle,
\end{align}
where the angular brackets represent an average over different initial conditions and disorder realizations. The structure factor, which is the Fourier transform of $C(r,t)$, is given by:
\begin{align}
\label{Eqn7}
S(k, t)=\int d\vec{r} e^{i \vec{k} \cdot \vec{r}} C(\vec{r}, t),
\end{align}
at a wave vector $\vec k$. The OJK theory studies the domain kinetics in a disorder free ferromagnet via defect dynamics~\cite{pw09}, which predict the functional form of the the equal-time correlation function as
\begin{align}
\label{OJK}
C_{\rm OJK}(r, t)=\frac{2}{\pi}\sin^{-1}({e^{-r^2/L^2}}),~~~~~~~~~~L(t)\sim t^{1/2}.
\end{align}
Fig.~\ref{Fig_4}(a) shows a scaling plot of the correlation function $C(r,t)$ versus $r/L$ for $\rho=0.01$ and $h_{\rm max}=0.5$ at four different times, as indicated. The numerical data clearly collapse onto a single curve, demonstrating dynamical scaling. This suggests that domain growth is characterized by a single length scale, and the morphology of the domains is scale invariant. A soild curve depicts the OJK function $C_{\rm OJK}(r, t)$, as defined in Eq.-\eqref{OJK}. Likewise, $S(k,t)$ also exhibits dynamical scaling, as illustrated in Fig.~\ref{Fig_4}(b). The black solid curve represents the Fourier transform of the OJK function in $d=2$. Furthermore, we observe that in the limit $k\rightarrow\infty$, $S(k,t)$ decays as $k^{-3}$, consistent with the  \textit{Porod's law}~\cite{b94,p82,op88} for the decay the structure-factor  $S(k,t)\sim k^{-(d+1)}$ at large $k$. As $d=2$, this yields the observed $k^{-3}$ behavior. This observation confirms the presence of sharp and well-defined boundaries between domains of differently oriented spins. These findings for $C(r,t)$ and $S(k,t)$ remain consistent across other values of $\rho$ and $h_{\rm max}$. Next, we compare the scaling functions with $C_{\rm OJK}(r, t)$ and its Fourier transform, represented by the solid black lines in Fig.~\ref{Fig_4}. At small distances, $C(r, t)$ aligns closely with $C_{\rm OJK}(r, t)$ but shows deviations at larger distances. Similar behavior is reflected in $S(k,t)$, where there are discrepancies at small $k$ values and good agreement at large $k$ values. These results differ from previous findings in RFIM, where a cusp-like behavior in $C(r,t)$ at small distances and a non-Porod-like decay of the $S(k,t)$ for large-$k$ were observed~\cite{sk14,kbp17}.

\begin{figure}
    \centering
    \includegraphics[width=0.69\textwidth]{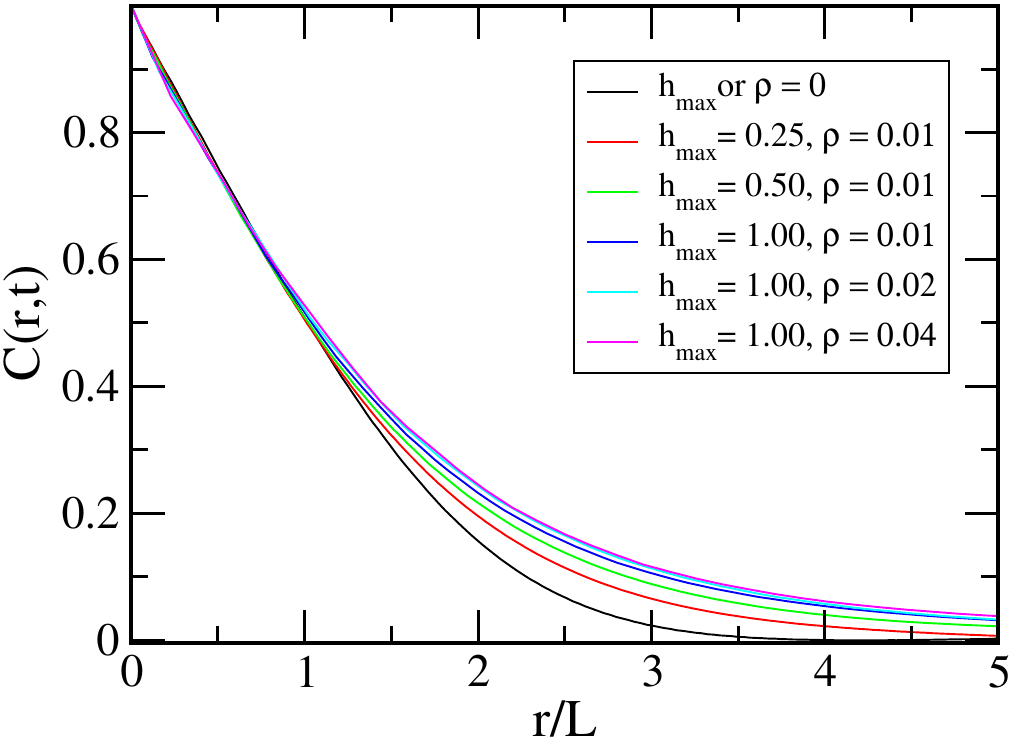}
    \caption{Scaling correlation function $C(r,t)$ versus $r/L$ at a fixed $t=10,000$ but for 
    different values of  $h_{\rm max}$ and $\rho$, as marked. The data for $h_{\rm max}$ or $\rho=0$ refers to the pure case.}
    \label{Fig_5}
\end{figure}
To quantify the impact of the impurity density $\rho$ and maximum field strength $h_{\rm max}$  on domain morphologies, we plot the scaled correlation functions $C(r, t)$ versus $r/L$ at a fixed time $t = 10,000$ but for different values of  $h_{\rm max}$ and $\rho$, as shown in Fig.~\ref{Fig_5}. A pertinent question arises: do the scaling functions depend explicitly on $\rho$ or $h_{\rm max}$? If the scaling functions remain invariant except through the time-dependent length scale $L$, they are referred to as the superuniversal~(SU). Fig.~\ref{Fig_5} clearly shows a distinct scaling forms of $C(r,t)$ {\it vs.} $r/L$ for different $h_{\rm max}$ and $\rho$, demonstrating a violation of SU, in agreement to the previous studies~\cite{CLMPZ12,kbp17,kc20}.
We also examined the SU of $S(k,t)$ (not shown here) across various values of $h_{\rm max}$ and $\rho$. Our findings indicate a consistent violation of SU due to deviations of scaling form of $S(k,t)$ at small wave vectors $k$. However, the tails of the structure factor consistently follow the Porod law, $S(k,t) \sim k^{-3}$ in $d=2$, in all cases studied.

\begin{figure}
    \centering
    \includegraphics[width=0.69\textwidth]{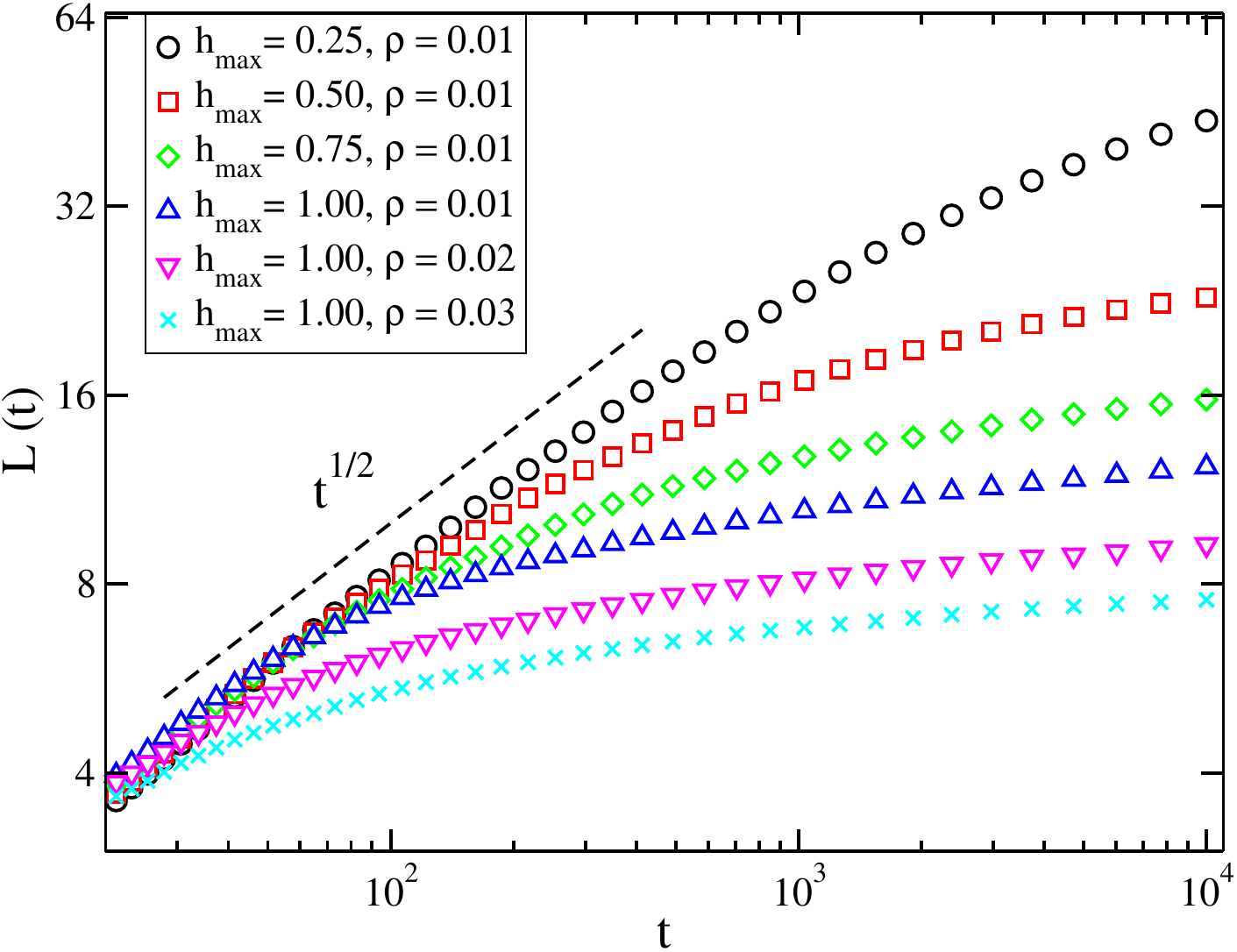}
    \caption{Log-log plot of the domain length scales $L(t)$ as a function of $t$  for different values of  $h_{\rm max}$ and $\rho$, as labeled. The dashed line represents the LAC law: $L(t)\sim t^{1/2}$.}
    \label{Fig_6}
\end{figure}

\begin{figure}
    \centering
    \includegraphics[width=0.99\textwidth]{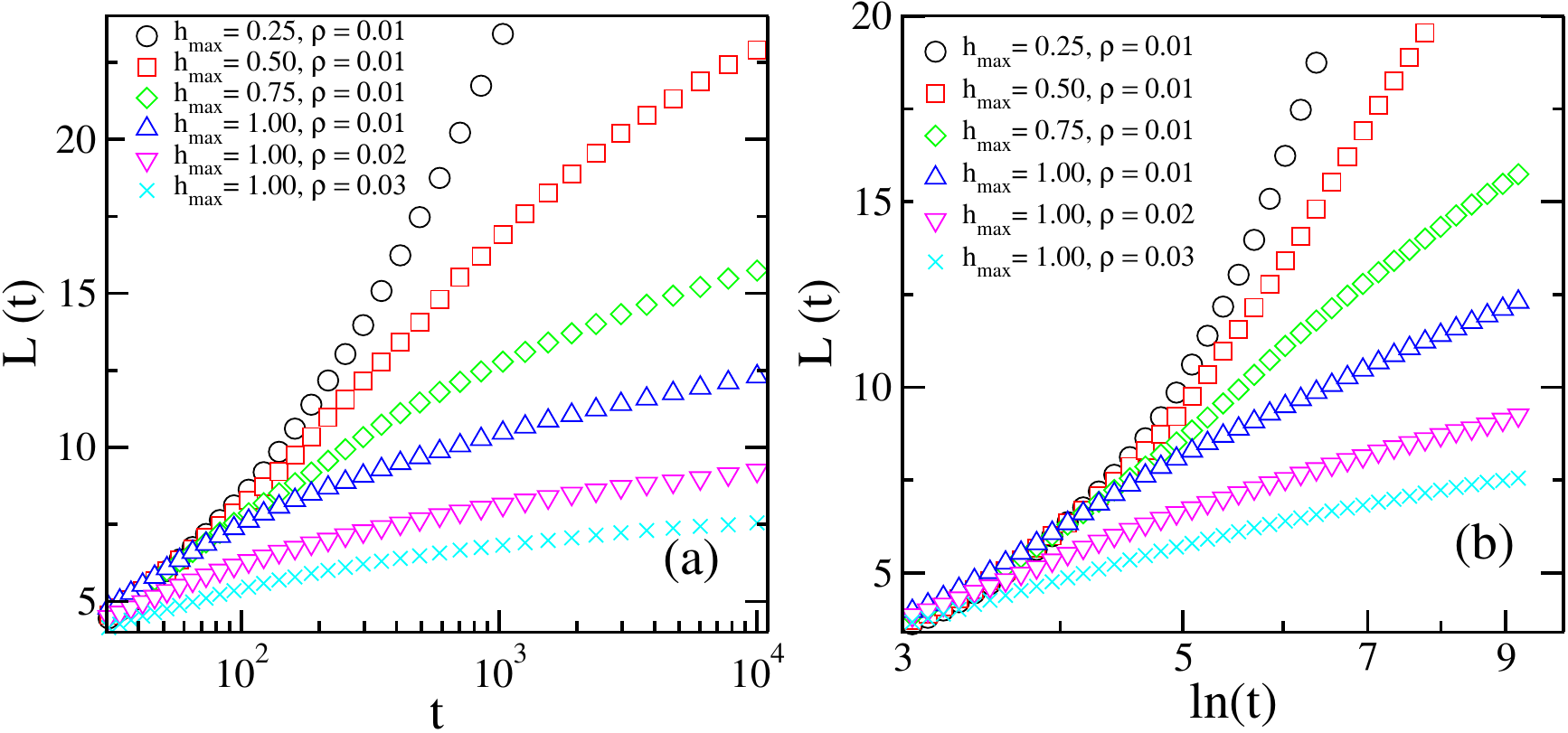}
    \caption{Semilog plots of (a) $L(t)$ against $t$ and (b) $L(t)$ against $\ln t$   different values of  $h_{\rm max}$ and $\rho$, as labeled.}
    \label{L_vs_lnt}
\end{figure}
We now move to investigate the domain growth laws. We define the average domain size $L(t )$ as the distance at which the correlation function $C(r,t)$ decreases to half of its maximum value, which is 1 at $r=0$. Fig.~\ref{Fig_6} shows the time dependence of $L(t )$ on a log-log scale. During the early stages of evolution, $L(t )$ exhibits power-law growth following the LAC law: $L(t)\sim t^{1/2}$. This indicates that, initially, the domain growth is not hindered by the disorder field. As time progresses, the growth process slows down due to the pinning of domains caused by the disorder. To check if the growth process is logarithmic, we plot $L(t)$ against $t$ on a log-linear scale in Fig.~\ref{L_vs_lnt}(a). We observe a logarithmic behavior of growth in an intermediate time regime, which can be seen crossing further to the growth slower than logarithmic at late times. This crossover occurs at lower $t$ with increase of $h_{\rm \max}$ or $\rho$.  In Fig.~\ref{L_vs_lnt}(b), we plot $L(t)$ with $\ln t$ on a log-linear scale. The data exhibits linearity in the late-$t$ regime, indicating a clear signature of an asymptotic \textit{double logarithmic growth} behavior, i.e., $\ln \ln t$. This deceleration of the domain growth process to a double logarithmic form manifests a significant influence of both parameters $\rho$ and $h_{\rm max}$ on the ordering dynamics of the model.

\begin{figure}
    \centering
    \includegraphics[width=0.99\textwidth]{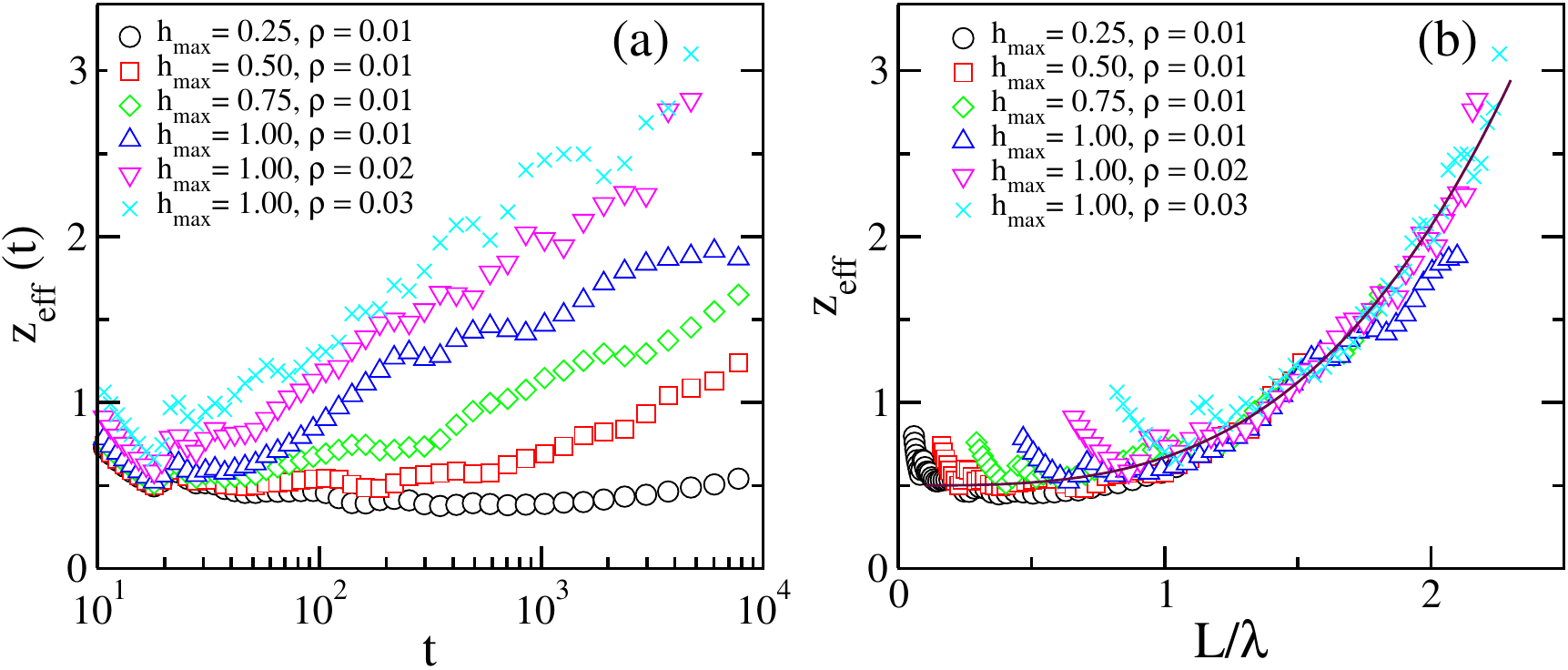}
    \caption{(a) Plot of $z_{\rm eff}(t)$, as defined in Eq.~\eqref{z_eff}, as a function of $t$ (on a log scale) for different values of  $h_{\rm max}$ and $\rho$, as labeled. (b) Scaling of $z_{\rm eff}$ with $L/\lambda$, where $\lambda$ is chosen to yield the best data collapse. A solid line is the fit to the form: $z_{\rm eff}=z_0+a(L/\lambda)^{\phi}$. The best fit gives $z_0=0.5$, $a\simeq 0.17$ and $\phi\simeq3.2$ with the reduced $\chi^2 \simeq 0.9$, which is quite good. }
    \label{zeff_vs_t}
\end{figure}

\begin{figure}
    \centering
    \includegraphics[width=0.6\textwidth]{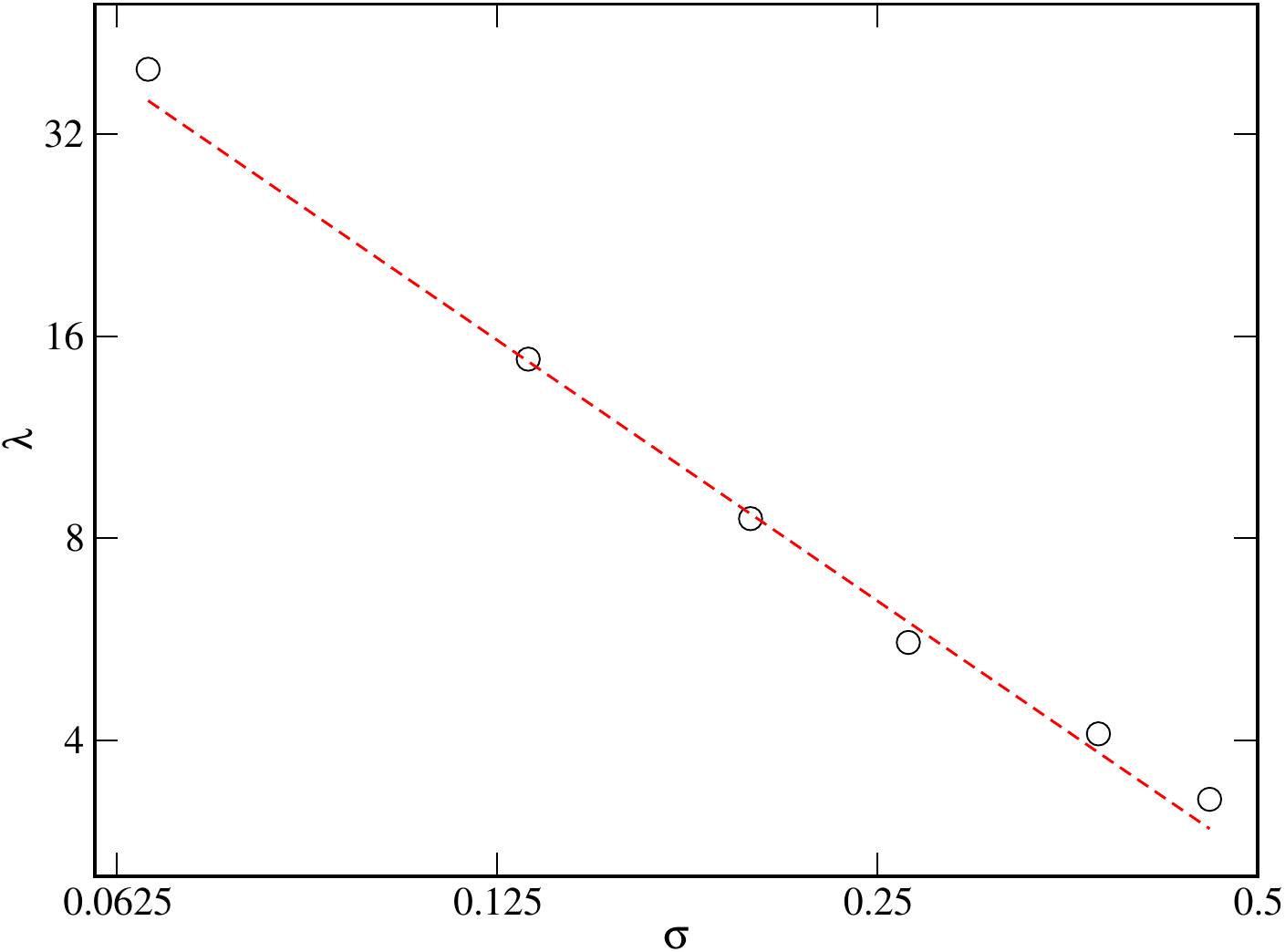}
    \caption{Plot of $\lambda$ vs $\sigma$ on a log-log scale. A red dashed line is the best fit to the power-law function: $\lambda\simeq1.08\sigma^{-1.3}$. }
    \label{crossover_lenth}
\end{figure}

To demonstrate a double logarithmic growth behavior, we define a key quantity, \textit{the effective exponent}, as
\begin{equation}
  \frac{1}{z_{\rm eff}}=\frac{d\ln L}{d(\ln \ln t)}. 
  \label{z_eff}
\end{equation}
We compute this quantity for the data shown in Fig.~\ref{Fig_6} and plot it as a function of $t$
in Fig.~\ref{zeff_vs_t}(a), note that $t$ is plotted on a log scale. Up to a sufficiently small time, say $t_0$, it is expected that $L(t)\sim t^{1/z}$ as the domains have not yet been affected by the disorder. This implies, according to the def.~\ref{z_eff}, $z_{\rm eff}=z/\ln t$. This is clearly visible in the Fig.~\ref{zeff_vs_t}(a), where until times $t\lesssim 20$, $z_{\rm eff} \sim 1/\ln t$.  For $t\gtrsim 20$, $z_{\rm eff}$'s develop a flat behavior, which corresponds a logarithmic behavior of $L(t)$. Notably, the time window in which $z_{\rm eff}$ is flat becomes smaller with increase of $h_{\rm \max}$ or $\rho$. Beyond this regime, $z_{\rm eff}$ shows a rise in the upward direction, indicating to a growth slower than logarithmic.  To identify this slow-growth regime, we collapse the $z_{\rm eff}$ data as a function of $L/\lambda$, where $\lambda$ is the length-scale for crossover from the logarithmic growth. As shown in Fig.~\ref{zeff_vs_t}(b), a nice data collapse is achieved by choosing $\lambda$ appropriately for different $h_{\rm max}$ and $\rho$. The collapsed data is fitted to the form $z_{\rm eff}(y) = z_0 + a y^\phi$. The best-fit parameters, represented by the solid line, are found to be $z_0 = 0.5$, $a \simeq 0.17$, and $\phi \simeq 3.2$. Then, using \eqref{z_eff},  we obtain the growth law in the crossover regime:
\begin{equation}
  \left[\frac{d\ln L}{d(\ln \ln t)}\right]^{-1}\simeq a\left(\frac{L}{\lambda}\right)^\phi \quad \Rightarrow ~L\sim\lambda\left[\left(\frac{\phi}{a}\right)\ln \ln t\right]^{1/\phi}. 
  \label{z_eff_form}
\end{equation}

Before the crossover, when $L<\lambda$, we have $z_{\rm eff}\simeq z_0$, implying a logarithmic growth law $L\sim (\ln t)^{1/z_0}$. With $z_0=0.5$ and $\phi=3.2$, our analysis shows that the domain growth exhibits a $(\ln t)^2$ behavior in the intermediate regime, which crosses over to a double-logarithmic form $(\ln \ln t)^{1/\phi}$ at late times. While the logarithmic-square growth ($\ln^2 t$) has been observed in earlier studies~\cite{pf85,ds85,a87,o94}, those and several others~\cite{oc90,pp93,v84,CLMPZ12} predicted a logarithmic growth law at late times. In contrast, our findings indicate an ultra slower, double-logarithmic growth. This behavior arises due to the accumulation of energy barriers associated with sites where the random field is centered at its maximum value, $h_{\rm \max}$. As the system evolves, these barriers become more prevalent, with their heights increasing exponentially with the characteristic domain-size. Assuming $E_B(L) \sim \exp(L^\phi)$ and the typical time to overcome such barriers scaling as $t \sim \exp(E_B/T)$, the domain growth is governed by a double-logarithmic law: $L \sim (\ln \ln t)^{1/\phi}$.

In Fig.\ref{crossover_lenth}, we plot the crossover length $\lambda$, which marks the onset of the double-logarithmic regime, as a function of $\sigma$, the width of the disorder distribution defined in Eq.~\eqref{sigma}. The data is presented on a double-logarithmic scale. A power-law fit, shown by the dashed line, yields $\lambda \simeq \sigma^{-1.3}$. The negative power-law exponent indicates that increasing disorder reduces the crossover length scale, signifying the relevance of disorder in slowing down of domain growth process.


\section{Summary and Outlook}\label{sec4}
Let us conclude this paper with a summary and discussion of our findings. We explored the domain growth kinetics in a disordered ferromagnet after a quench at a finite temperature under the influence of a disorder field. This disorder field originates from magnetic impurities in the system. We modeled the dynamics of the system using Monte Carlo simulations with Glauber spin-flip kinetics. We observed that, for a specified impurity density $\rho$ and disorder field strength $h_{\rm max}$, domain growth initially accelerates but decelerates at a later time due to pinning of domain walls at impurity sites. The average domain size $L$ at a given Monte Carlo time ($t$) decreases as one increases $h_{\rm max}$ increases while keeping $\rho$ constant, or vice-versa.

We characterize the morphology of domains by computing the equal-time correlation function $C(r,t)$ and the structure factor $S(k,t)$. The numerical data for scaled $C(r,t)$ across different $t$ at fixed $\rho$ and $h_{\rm max}$ exhibit data collapse, indicating dynamical scaling in the domain growth process as observed in many studies for domain growth on RFIM~\cite{accp08,rc293,skb11,o94,pp93,oc90}. Moreover, $C(r,t)$ at small distances is similar to $C_{\rm OJK}(r,t)$. In sharp contrast to RFIM, we do not see any cusp-like behavior in $C(r,t)$ at short distances. Similar to $C(r,t)$, $S(k,t)$ at different times also show dynamical scaling. For large-$k$, $S(k,t)$ obeys Porod's law: $S(k,t)\sim k^{-3}$ in a spatial dimension of $d=2$. This confirms that the interfaces between differently ordered domains are sharp. However, in RFIM, a non-Porod-like decay of $S(k,t)$ has been observed, indicating scattering from rough fractal interfaces. We then compared the dynamical scaling forms of $C(r,t)$ for different $h_{\rm max}$ and $\rho$, which showed a distinct scalig forms, indicating a violation of superuniversality~(SU), in agreement to previous studies on  RFIM~\cite{pp93,CLMPZ12,kbp17}.

Next, we studied the time dependence of $L(t)$, which we calculate from the decay of $C(r,t)$. At early stage of evolution, $L(t)$ follows Lifshitz-Allen-Cahn (LAC) law: $L(t)\sim t^{1/2}$. As time progresses, it transitions to a logarithmic growth, which eventually crosses to a \textit{double logarithmic growth law}, with the crossover time depending on both $\rho$ and $h_{\rm max}$. From the analysis of the effective exponent, we determine the asymptotic growth law as $L(t) \sim \lambda (\ln \ln t)^{1/\phi}$, where the growth-exponent is is found to be $\phi \simeq 3.2$ and the crossover length scales as $\lambda \sim \sigma^{-1.3}$.
These findings offer valuable insights into the nonequilibrium behavior in disordered ferromagnets, and we hope that our results will be applied to experimental findings,  particularly in systems doped with magnetic particles.

In recent years, understanding the ground state properties and superuniversality of the RFIM has garnered enormous attention~\cite{sk14,bb16,kbp17}. The correlation function $C(r,t)$ exhibits a cusp-like singularity, and the tail of the structure factor $S(k,t)$ demonstrates non-Porodian decay, which suggests scattering from fractal domains~\cite{sk14,kb17}. To explore these properties, advanced numerical techniques such as branch-and-cut programs~\cite{sdmr95}, matching algorithms~\cite{h11}, and a hybrid of genetic algorithms with cluster exact approximations~\cite{hprb99} have been employed. Looking ahead, we plan to investigate the ground state properties,  aging, and relaxation kinetics for a quench at $T=0$ for the model presented above, as quenches to zero and nonzero temperatures can exhibit distinct growth mechanisms. Additionally, we aim to extend these studies to higher dimensions,  where domain growth dynamics and the associated scaling functions are expected to differ appreciably.

\ \\
\noindent{\bf Acknowledgments:} SH acknowledges financial support from IISER Mohali through a Senior Research Fellowship. PD acknowledges financial support from SERB, India through a start-up research grant (SRG/2022/000105). MK acknowledges financial support from ANRF, India through a Ramanujan Fellowship (RJF/2022/000149).

\end{document}